\documentclass[aps,twocolumn,superscriptaddress,prb,amsmath]{revtex4-2}
\usepackage{multirow}
\usepackage[latin9]{inputenc}
\usepackage{epstopdf}
\usepackage{mathrsfs}
\usepackage{txfonts}
\usepackage{amssymb}
\usepackage{graphicx,subfigure,float}% Include figure files
\usepackage{dcolumn}% Align table columns on decimal point
\usepackage{bbm}% bold math
\usepackage{bm}
\usepackage{color}
\usepackage[colorlinks, linkcolor=blue, citecolor=blue, urlcolor=blue]{hyperref}
\usepackage{lipsum}
\usepackage{gensymb}
\usepackage{amsmath}
\usepackage{caption}
\usepackage{bm}
\begin{document}
%++++++++++++++++++++++++++++++++++++++++++++++++++++
% Title of the article
\title{High-Throughput Discovery of Two-Dimensional Materials Exhibiting Strong Rashba-Edelstein effect}
%++++++++++++++++++++++++++++++++++++++++++++++++++++
\author{Binchang Zhou}
\affiliation{Key Laboratory of Low Dimensional Materials and Application Technology of Ministry of Education, School of Materials Science and Engineering, Xiangtan University, Xiangtan 411105, China}
\author{Baoru Pan}
\affiliation{Key Laboratory of Low Dimensional Materials and Application Technology of Ministry of Education, School of Materials Science and Engineering, Xiangtan University, Xiangtan 411105, China}
\affiliation{Hunan Provincial Key laboratory of Thin Film Materials and Devices, School of Materials Science and Engineering, Xiangtan University, Xiangtan 411105, China}
\author{Pan Zhou}
\email{zhoupan71234@xtu.edu.cn}
\affiliation{Hunan Provincial Key laboratory of Thin Film Materials and Devices, School of Materials Science and Engineering, Xiangtan University, Xiangtan 411105, China}
\author{Yuzhong Hu}
\affiliation{Hunan Provincial Key laboratory of Thin Film Materials and Devices, School of Materials Science and Engineering, Xiangtan University, Xiangtan 411105, China}
\author{Songmin Liu}
\affiliation{Hunan Provincial Key laboratory of Thin Film Materials and Devices, School of Materials Science and Engineering, Xiangtan University, Xiangtan 411105, China}
\author{Lizhong Sun}
\email{lzsun@xtu.edu.cn}
\affiliation{Key Laboratory of Low Dimensional Materials and Application Technology of Ministry of Education, School of Materials Science and Engineering, Xiangtan University, Xiangtan 411105, China}
%++++++++++++++++++++++++++++++++++++++++++++++++++++

%============================================
\date{\today}
%============================================
\begin{abstract}
The Rashba-Edelstein effect (REE), which generates spin accumulation under an applied electric current, quantifies charge-to-spin conversion (CSC) efficiency in non-centrosymmetric systems. However, systematic investigations of REE in two-dimensional (2D) materials remain scarce. To address this gap, we perform a comprehensive symmetry analysis based on the 80 crystallographic layer groups, elucidating the relationship between materials' symmetries and the geometric characteristics of the REE response tensor. Our analysis identifies 13 distinct symmetry classes for the tensor and reveals all potential material candidates. Considering the requirement of strong spin-orbit coupling for a large REE response, we screen the C2DB database and identify 54 promising 2D materials. First-principles calculations demonstrate that the largest REE response coefficients in these materials exceed those reported for other 2D systems by an order of magnitude, indicating exceptionally high CSC efficiency. Focusing on three representative materials, including HgI$_2$, AgTlP$_2$Se$_6$ and BrGaTe, we show that their large response coefficients can be well explained by effective $\bm{k \cdot p}$ models and the characteristic spin textures around high-symmetry points in momentum space. This work provides a systematic framework and identifies high-performance candidates, paving the way for future exploration of REE-driven CSC in 2D materials.
\end{abstract}
\maketitle
%============================================
\section{Introduction}
\indent The Rashba-Edelstein effect (REE) refers to the generation of a non-equilibrium spin accumulation in response to an applied electric field, arising from spin-momentum locking in systems with broken inversion symmetry and significant spin-orbit coupling\cite{rashba-effect,edelstein-effect,REE1,REE2}. This phenomenon originates from an asymmetric redistribution of carriers in momentum space, where the spin texture (ST) determined by the underlying crystal symmetry and electronic structure translates the momentum-space imbalance into a net spin polarization\cite{book1}. In general crystalline environments, the direction and magnitude of the induced spin accumulation depend sensitively on the symmetry of the spin-orbit field and the anisotropy of the band structure. The Rashba-induced spin polarization enables efficient charge-to-spin conversion, underpinning a range of spintronic phenomena and applications, such as current-driven magnetization switching\cite{mag-switch1,mag-switch2}, spin-orbit torques\cite{SOT1,SOT2,mag-switch2}, spin-charge interconversion at interfaces\cite{interfaces1,interfaces2,interfaces3}, and the realization of nonreciprocal transport effects\cite{nonreciprocal1,nonreciprocal2}. The Rashba coupling, which can be tuned through interfacial engineering or external stimuli such as ultrafast optical control and femtosecond light pulses, enables dynamic manipulation of spin polarization and offers great potential for low-power, ultrafast spintronic devices\cite{devices1,devices2,devices3,devices4,devices5,devices6,devices7}.\\
%============================================
\indent While the microscopic mechanism of REE has been extensively studied in model Hamiltonians and idealized heterostructures\cite{rashba-effect,edelstein-effect,heterostructures1,heterostructures2,heterostructures3,model1,mera2021different}, its symmetry properties in realistic two-dimensional (2D) materials remain largely unexplored. In particular, the symmetry analysis of the REE tensor for 3D system has been studied\cite{analogs}, but a systematic group-theoretical analysis of the REE tensor for low-symmetry 2D materials is still lacking. On the materials side, only a limited number of 2D systems with large REE conversion coefficients have been reported, mostly based on prototypical interfaces\cite{interfaces1,interfaces2,interfaces3} or heavy-element compounds\cite{mag-switch2,heavy2}. Comprehensive computational searches for intrinsic 2D materials with strong REE responses have yet to be performed. This gap constrains the discovery and application of efficient spin-charge interconversion in 2D materials. Motivated by recent progress in spintronics devices\cite{mag-switch1,recent-progress2,recent-progress3,recent-progress4} and rapidly expanding 2D material databases\cite{C2DB}, a combined symmetry analysis and high-throughput computational screening are needed to identify promising candidates and advance the development of REE-based spintronic devices.\\
%============================================
\indent In this work, we begin by employing group-theoretical analysis to classify the symmetry properties of the REE axial tensor in 2D materials, based on their layer group (LG) symmetries. To identify systems with potentially strong REE response tensors, we conduct a targeted screening of the 2D materials database and select 54 candidate compounds with distinct symmetry characteristics. First-principles calculations reveal that these systems exhibit REE tensor components with magnitudes surpassing those reported in previously studied materials, underscoring their potential for realizing substantial current-induced spin accumulation. To further uncover the microscopic mechanisms underlying these large REE responses, we construct symmetry-constrained $\bm{k \cdot p}$ models grounded in the relevant irreducible representations for three representative materials, which faithfully capture the symmetry-allowed formation of STs.\\
%============================================
\section{Symmetry Analysis of the Rashba-Edelstein Tensor in Layer group}
\indent In the framework of Boltzmann transport theory, the equilibrium distribution of electrons in a crystal is described by the Fermi distribution function, denoted as \( f_{\bm{k}}^0 \). In the equilibrium state, time-reversal symmetry plays a crucial role. Specifically, due to this symmetry, the expectation values of the spin operator \( \bm{S} \) at momenta \( \bm{k} \) and \( -\bm{k} \) are equal in magnitude but opposite in direction. As a result, their contributions cancel each other, leading to a vanishing net spin polarization. However, when a charge current is induced, an external electric field perturbs the electron system, driving it out of equilibrium~\cite{equilibrium1,equilibrium2}. The Fermi surface undergoes a shift, and the electron distribution deviates from its equilibrium form, evolving into a non-equilibrium distribution described by \( f_{\bm{k}} = f_{\bm{k}}^0 + \delta f_{\bm{k}} \). This deviation gives rise to a finite net spin accumulation. The resulting spin accumulation \( \delta \bm{s} \) can be quantified by the expression:
\begin{equation}
\delta \bm{s} = \sum_{\bm{k}} \langle \bm{S} \rangle_{\bm{k}} \delta f_{\bm{k}},
\label{eq:entropy_change}
\end{equation}
This expression involves a summation over all momenta \( \bm{k} \) in the Brillouin zone, where \( \langle \bm{S} \rangle_{\bm{k}} \) denotes the expectation value of the spin operator at momentum \( \bm{k} \), and \( \delta f_{\bm{k}} \) represents the deviation of the distribution function from its equilibrium form. Notably, the charge current density can also be expressed in terms of the same non-equilibrium correction \( \delta f_{\bm{k}} \)~\cite{ziman1972principles}:
\begin{equation}
\bm{j_c} = -\frac{e}{V} \sum_{\bm{k}} v_{\bm{k}} \delta f_{\bm{k}}.
\label{eq:current}
\end{equation}
Then the Rashba-Edelstein response tensor $\chi^{ij}$ is defined as a ratio of the spin accumulation $\delta \bm{s}$ and the charge current $\bm{j_c}$\cite{roy2022long}. The first subscript of the REE tensor denotes the polarization direction of the electron spin, while the second subscript represents the direction of the applied electric field. Both of the two indices are independent. It describes the conversion efficiency and direction from charge current to spin accumulation.
%============================================
\indent The REE tensor is an axial tensor, and a system would exhibit a nonzero response if it do not possess spatial inversion symmetry. Moreover, the REE tensor is also constrained by other crystal symmetry\cite{analogs}. If a symmetry operation is represented by a 3$\times$3 matrix $R$, its transformation relation should be
\begin{equation}
\chi'_{ij} = (\det R) \cdot \sum_{k,l} R_{ik} R_{jl} \chi_{kl},
\label{eq:tensor_transform}
\end{equation}
where $\det$ denotes the determinant of the matrix and $i, j, k, l \in \{x, y, z\}$\cite{book2}. If the system remains invariant under this symmetry operation, it is required that
\begin{equation}
\chi'_{ij} = \chi_{ij}.
\label{eq:tensor_invariant}
\end{equation}
%============================================
\indent When describing 2D materials, we use LGs rather than point groups, as the latter can introduce ambiguity regarding the orientation of the symmetry axis. For example, a 2D system with $C_2$ symmetry, which is a polar point group, can have a two-fold rotation axis that either lies within or is perpendicular to the 2D plane. These two configurations lead to distinct symmetry constraints.\\
%============================================
\indent To systematically derive the allowed forms of the REE tensor $\chi$ across all 80 LGs, the symmetry operations of a single layer system are classified into seven distinct classes. This classification is based on the similarity of constraints that each operation imposes on the REE tensor $\chi$. Details of the seven distinct classes symmetry operations are provided in Sec. I of Supplemental Material (SM)\cite{sup}. Then we analyze how the generators of each LG constrain the tensor structure\cite{bilbao1, bilbao2}. It is important to note that pure translation operations do not impose any restrictions on the REE tensor\cite{analogs}. By systematically deriving the symmetry constraints imposed by all relevant operations, we obtain the general form of the REE tensor for each LG. A detailed example of the derivation process for the REE tensor for LG $p112$ (No. 3) is provided in Sec. I of the SM. The results for all LGs are compiled in Sec. II and Sec. III of the SM, offering a comprehensive and practical reference for determining the tensor structure in different symmetry environments. Furthermore, due to the non-periodic nature of the vacuum layer in 2D materials, applying an external electric field perpendicular to the plane (along the vacuum direction) does not generate spin-polarized currents. Therefore, the last column of the tensor is zero for all 2D materials, and we omit this column for them.\\
\begin{figure}
	\centering
	\includegraphics[trim={0.0in 0.0in 0.0in 0.0in},clip,width=0.75\linewidth]{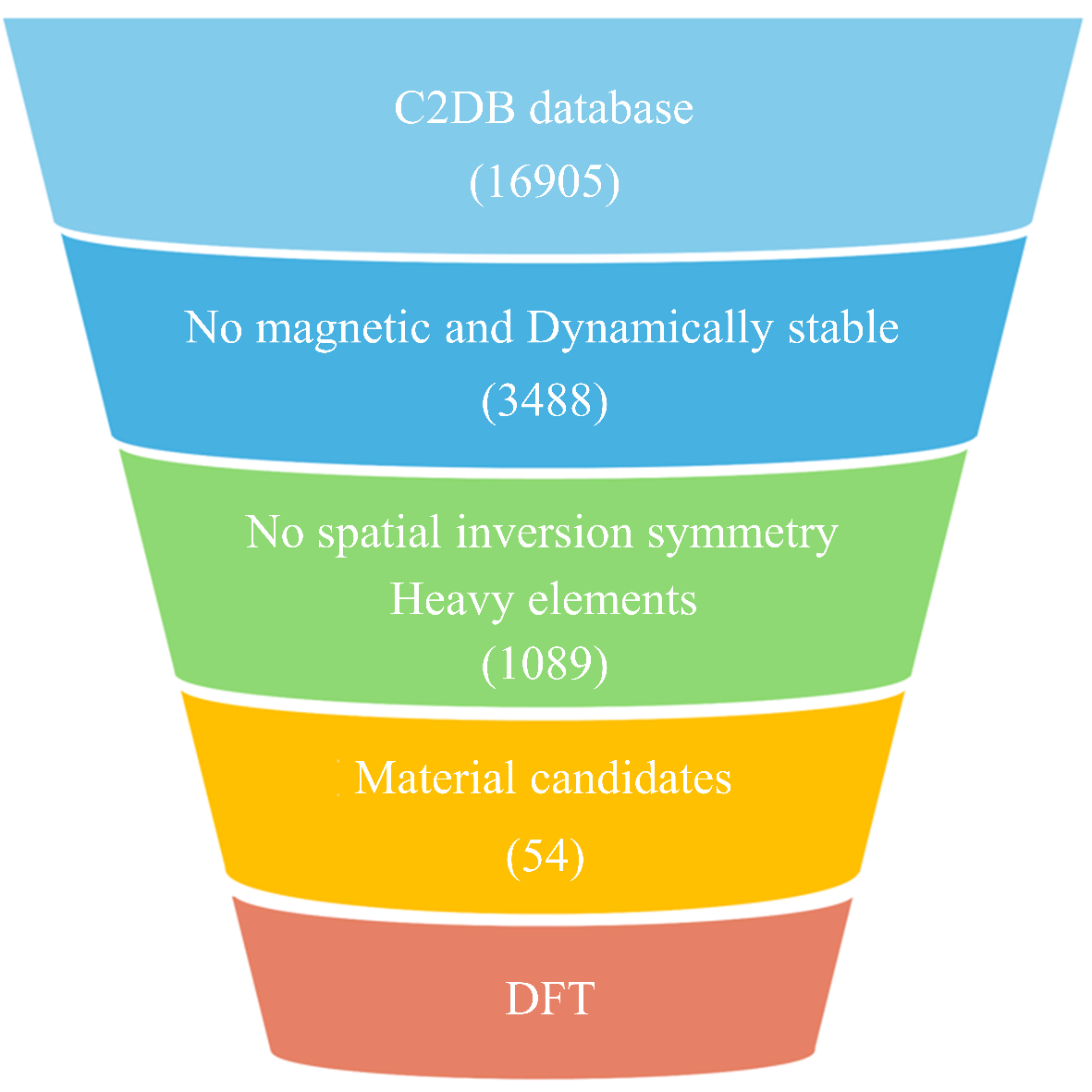}
    \captionsetup{labelsep=period}
	\caption{The workflow of high-throughput screening process.}\label{fig1}
\end{figure}
\begin{figure*}
    \centering
    \includegraphics[trim={0.0in 0.0in 0.0in 0.0in},clip, width=\textwidth]{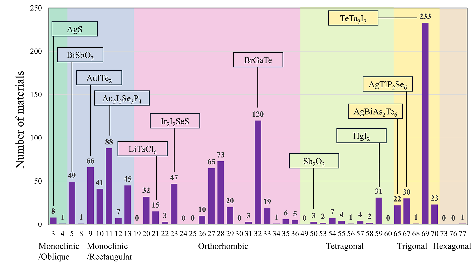}
    \captionsetup{labelsep=period}
    \caption{Distribution of materials according to LG with inversion symmetry broken. The background color coding refers to the crystal system (the monoclinic crystal system is further classified according to its lattice type). Part of the candidate materials from different categories are shown in the figure.}
    \label{fig2}
\end{figure*}
%============================================
\section{Rules for Material Screening and Selection}
%============================================
\indent The process of identifying materials with large REE tensor is outlined in Fig. 1. Initially, we examine the C2DB database, which contains a total of 16,905 2D materials, with energy above hull values ranging from 0 to 0.2 eV/atom. From this pool, we select non-magnetic and dynamically stable materials, totaling 3,488 candidates. It is worth mentioning that the dynamical stability in this database is determined solely by the non-negative phonon frequencies at high-symmetry points. For experimental synthesis of the materials, it is necessary to further investigate the phonon dispersion at other k points. As significant SOC typically correlates with large REE, we further narrow down the selection to materials that contain elements from the fourth to the sixth period and exhibit broken spatial inversion symmetry. This subset consists of 1,089 materials. Next, we classify these 1,089 materials based on their LGs, as shown in Fig. 2. The LGs of the 2D materials are sourced from the C2DB, which utilizes the SPGLIB library for symmetry determination. To confirm the symmetries of all studied materials, we further validated the results using SPGLIB with a distance tolerance of 0.01 {\AA} for atomic positions and an angle tolerance of 1$^\circ$ for lattice vectors. We observe that the distribution of materials among the LGs is highly uneven. Among them, the LG with the most materials has 233 materials, and there are 8 LGs with no materials. We further categorize the LGs according to their REE tensor forms, identifying 13 distinct forms across 80 LGs, as detailed in SM, Tab. S2.
%============================================
\begin{figure*}
    \centering
    \includegraphics[trim={0.0in 0.0in 0.0in 0.0in},clip, width=\textwidth]{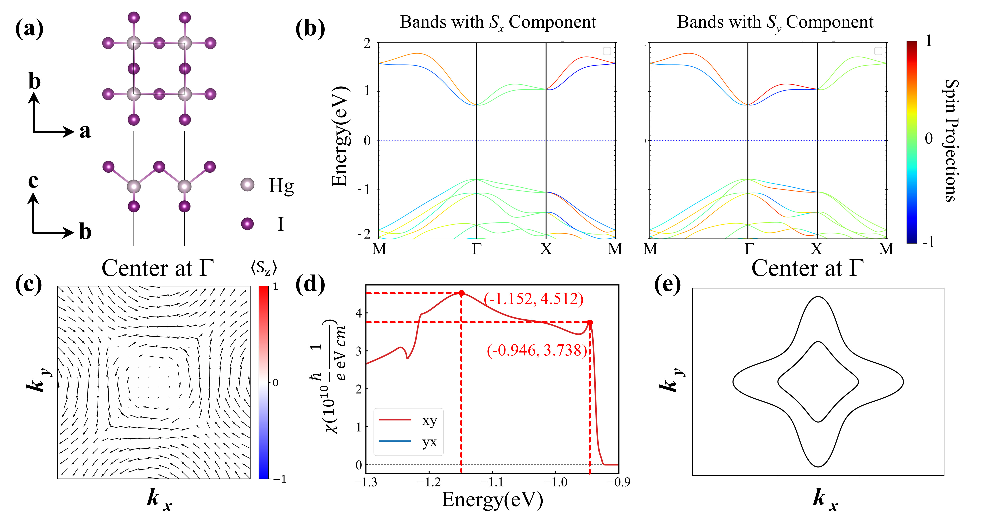}
    \captionsetup{labelsep=period}
    \caption{Structure and electronic properties of the HgI$_2$. (a) Top view and side view of 2D HgI$_2$. (b) Spin-polarized bands for $S_x$, $S_y$ spin projections of HgI$_2$ shown along the $M$ - $\Gamma$ - $X$ - $\Gamma$ path. (c) In-plane ST around $\Gamma$. ($S_x$, $S_y$) components are represented by the arrows. $S_z$ component is represented by the color.(d) Calculated $\chi_{xy}$ and $\chi_{yx}$ as a function of the chemical potential. Due to symmetry considerations, the component $\chi_{xy}$ is equivalent to $\chi_{yx}$. (e) The constant energy contour for $E = -0.946$ eV.}
    \label{fig3}
\end{figure*}
%============================================
Our subsequent material selection will be based on these 13 categories. We will select an appropriate number of representative materials from each category for detailed calculation and analysis. Of course, since the number of materials varies significantly across different categories, our general strategy is to select more representative materials for categories with a larger number of entries. Moreover, since many materials share highly similar crystal structures and elemental compositions from the same group, we select only a few representative materials for each structure. For example, in LG $p3m1$ (No. 69), these 233 materials ultimately correspond to 38 distinct structures. Each structure corresponds to a set of materials where the specific atoms occupying the same Wyckoff positions belong to the same main group, while the materials still maintain similar electronic structures and properties. In contrast, materials across different sets exhibit structural distinctions and consequently divergent electronic characteristics. Among these 38 sets, the largest contains 50 materials, while the smallest consists of only 1. To ensure synthesis feasibility and broaden discovery potential, we focus on sets containing at least 3 materials, resulting in 14 candidate sets. Since strong Rashba coupling is generally associated with heavy elements possessing substantial SOC\cite{largeSOC}, we prioritize materials with higher atomic numbers within each set. From each of the 14 sets, we select the material with the largest atomic number as a representative, yielding 14 candidates. Notably, structures with 20 or more atoms per unit cell were omitted, as they often imply low synthetic accessibility and high structural complexity.  After first-principles calculations, we selected materials exhibiting a REE value exceeding those of the In$_2$Se$_3$ and WTe$_2$ bilayers\cite{analogs,ferroWTe2}, identifying 9 promising candidates in this category. After applying the same screening approach and calculations for the other LGs within this category, resulting in 11 representative materials for this category. We then extend this method to all other categories and ultimately select 54 representative materials for REE tensor computation, as summarized in Supplementary Table S3. In the main manuscript, we focus on 3 of these materials for a detailed discussion of their ST and the origins of their large REE tensors. For completeness, we selected 9 materials that belong to different REE tensor forms to briefly illustrate the symmetry properties, as the relevant results for other materials can be obtained using a similar method. The remaining 9 materials can be classified based on their LGs and the resulting response tensor. First, AgS, Sb$_2$O$_3$, and AgBiAs$_2$Te$_6$ exhibit an identical fundamental form of response tensor, while their non-zero element relationships differ significantly. This distinction originates from the additional generator operations (e.g., C$_{nz}$ or S$_{nz}$), which impose specific constraints on the non-zero elements. In contrast, BiSbO$_3$, AuITe$_2$, Au$_2$I$_2$Se$_3$P$_4$, and LiTaCl$_6$ display different basic forms of response tensor. This difference is directly attributed to variations in their LGs (e.g. M$_z$, C$_{2x}$, M$_x$ and C$_{2z}$), thereby leading to fundamental discrepancies. Finally, Ir$_2$I$_2$SeS and TeTaI$_7$ retain an identical form of the response tensor, yet exhibit disparities in the relationships among their non-zero elements. This is attributed to the fact that mirror operations impose symmetry constraints that result in the diagonal elements of the response tensor being zero.\\
%============================================
\begin{figure*}
    \centering
    \includegraphics[trim={0.0in 0.0in 0.0in 0.0in},clip, width=\textwidth]{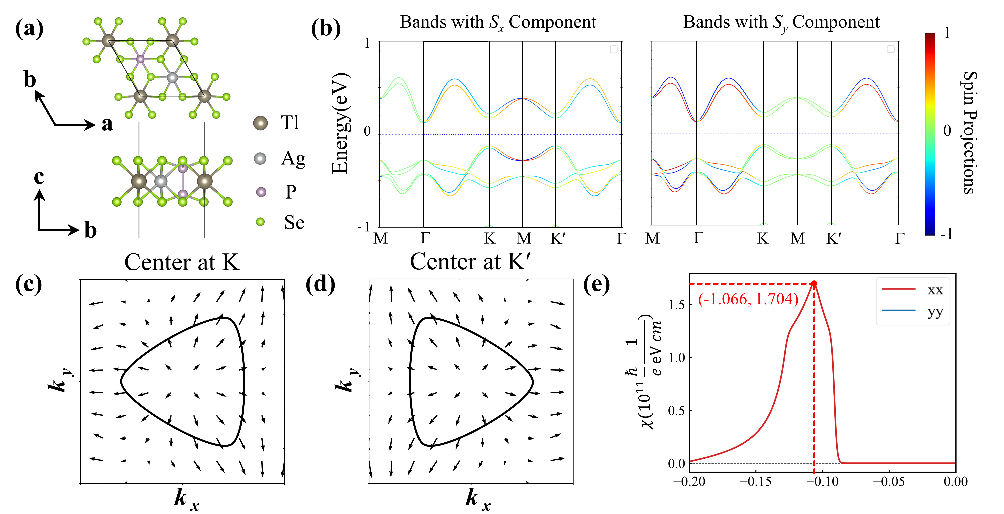}
    \captionsetup{labelsep=period}
    \caption{Computational analysis of monolayer AgTlP$_2$Se$_6$. (a) Top and side view of the crystal structure. (b) Spin-polarized energy bands with $S_x$, $S_y$ spin projections, plotted along the high-symmetry path. (c)-(d) The contour with the superimposed in-plane ST around $K$ and $K'$. ($S_x$, $S_y$) components are symbolized by the arrows. $S_z$ components are omitted. (e) The magnitude of the $\chi$ tensor is calculated as a function of the chemical potential. Symmetry analysis dictates that the component $\chi_{xx}$ is equal to $\chi_{yy}$.}
    \label{fig4}
\end{figure*}
%============================================
\section{Representative material H\lowercase{g}I$_2$}
%============================================
\indent The crystal structures of HgI$_2$ are shown in Fig. 3(a), which consist of three atomic layers with the middle Hg layer sandwiched between two I monolayers. It belongs to $P\overline{4}m2$ layer group (No. 59), with its point group $D_{2d}$. Since both Hg and I are heavy elements, their SOC effects are significant. The band structure with SOC is shown in Fig. 3(b). It is observed that the bands exhibit spin splitting, except at the time-reversal invariant points ($\Gamma$, M, and X), with both the valence band top and conduction band bottom located near the $\Gamma$ point. The wave vector point groups of $\Gamma$ is equal to its crystallographic point group $D_{2d}$. According to the definition of Mera Acosta\cite{mera2021different}, this point group is nonpolar nonchiral wave vector point group and can produce a kind of Dresselhaus ST around the $\Gamma$ point. Concretely, the irreducible representation of double degenerate bands at $\Gamma$ point is $\Gamma_6$. According to the invariant theory of $\bm{k \cdot p}$ model, the effective Hamiltonian of it can be write as $H^\Gamma_{\text{split}} = k_y \sigma_x + k_x \sigma_y$, which would make the two bands form tangetial-radial ST. The ST of HgI$_2$ from first-principles calculations are shown in Fig. 3(c). It is revealed that its spin direction is parallel to the wave vector along the diagonal, and perpendicular to the wave vector in the $x$ and $y$ directions, which is consistent with the analysis results of the $\bm{k \cdot p}$ model. \\
%============================================
\indent Based on the foundational theory outlined in Sec. II and the comprehensive derivation detailed in Sec. I to III of SM\cite{sup}, 2D HgI$_2$ has only two nonzero REE tensor components, $\chi_{xy} = \chi_{yx}$. Figure 3(d) shows the magnitude of these components as a function of the chemical potential $E$. We find that the values remain large over a wide energy range below the Fermi level, reaching extreme values at -1.152 eV and -0.946 eV, with values of $4.512 \times 10^{10}\,\frac{\hbar}{e} \frac{1}{\text{eV} \cdot \text{cm}}$ and $3.738 \times 10^{10}\,\frac{\hbar}{e} \frac{1}{\text{eV} \cdot \text{cm}}$, respectively. These values are larger than the maximal values of other typical 2D materials, such as In$_2$Se$_3$ ($1.31 \times 10^{10}\,\frac{\hbar}{e} \frac{1}{\text{eV} \cdot \text{cm}})$\cite{analogs} and WTe$_2$ ($1.4\times 10^{10}\,\frac{\hbar}{e} \frac{1}{\text{eV} \cdot \text{cm}}$)\cite{ferroWTe2}. The spin distribution at the energy of -0.946 eV [integrating Figs. 3(c) and 3(e) for a comprehensive view] reveals that, when the Fermi surface shifts along the $x$ ($y$) direction, the absence of $-y$ ($-x$) directions in the net spin accumulation prevents a reduction in spin accumulation, which results in the system hosting a large value of the REE tensor. By comparing existing literature\cite{analogs,ferroWTe2}, we found that this is the first time we have discovered such a large REE tensor formed by the Dresselhaus ST.\\
%============================================
\begin{figure*}
    \centering
    \includegraphics[trim={0.0in 0.0in 0.0in 0.0in},clip, width=\textwidth]{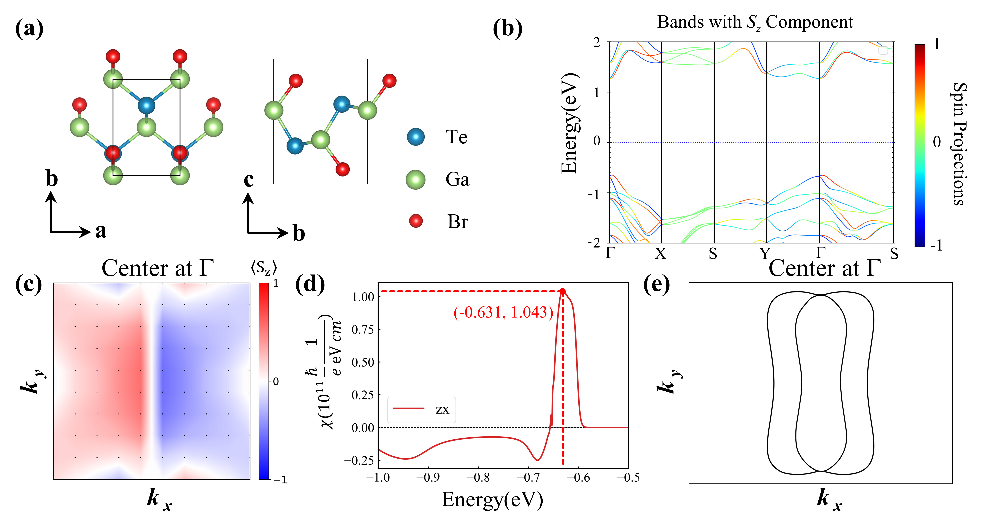}
    \captionsetup{labelsep=period}
    \caption{The Structural and electronic characteristics of BrGaTe. (a) Top view and side view of the crystal structure. (b) Spin-polarized bands for $S_z$ spin projection calculated along the high symmetry path. (c) In-plane ST around $\Gamma$. ($S_x$, $S_y$) components are represented by the arrows and the normal component $S_z$ by the color. (d) Calculated magnitude of the REE as a function of the chemical potential. The only component that exists is $\chi_{zx}$. (e) The constant energy contour for $E = -0.631$ eV.
}
    \label{fig5}
\end{figure*}
%============================================
\section{Representative material A\lowercase{g}T\lowercase{l}P$_2$S\lowercase{e}$_6$}
%============================================
\indent Monolayer AgTlP$_2$Se$_6$ exhibits a trigonal lattice structure, as depicted in Fig. 4(a), belonging to the LG $P312$ (No. 67) with a corresponding point group $D_3$. The structure comprises five atomic layers, with Ag and Tl atoms situated in the same plane, while the other two atom types are mirror-symmetrically distributed on the upper and lower sides. The electronic band structure, incorporating SOC, is shown in Fig. 4(b). A notable Zeeman-type spin splitting is observed near the $K$ and $K'$ points around the Fermi level, with an indirect band gap. The valence band maximum (VBM) is located at the $K$ and $K'$ points, while the conduction band minimum (CBM) resides at the $\Gamma$ point. Given the critical role of the VBM in CSC efficiency, the STs of the VBM bands are illustrated in Figs. 4(c) and 4(d). At these $K$ and $K'$ points, the wave vector point groups are $C_3$, corresponding to the Rashba-Weyl ST \cite{mera2021different}. To elucidate the mechanisms driving the formation of these STs, we calculate the irreducible representations at the $K$ and $K'$ points, specifically $K_4$ and $K_6$. This allows us to derive the $\bm{k \cdot p}$ spin-splitting Hamiltonians at these points, expressed as:
\[
H^K_{\text{split}} = -(k_x - k_x^2 + k_y^2)\sigma_x - (2k_x k_y + k_y)\sigma_y - (k_x^2 + k_y^2)\sigma_z
\]
\[
H^{K'}_{\text{split}} = -(k_x + k_x^2 - k_y^2)\sigma_x + (2k_x k_y - k_y)\sigma_y + (k_x^2 + k_y^2)\sigma_z
\]
Both Hamiltonians are expanded up to second-order $k$-polynomials to accurately capture the spin textures around these points. The $S_z$ spin accumulation is fully compensated by the term $(k_x^2 + k_y^2)\sigma_z$, and constant terms are omitted as they do not contribute to the spin texture.\\
%============================================
\indent Symmetry analysis of the REE tensor indicates that, due to inherent symmetry constraints, only the diagonal components $\chi_{xx}$ and $\chi_{yy}$ can be nonzero, while all other tensor elements vanish. Figure 4(e) presents the calculated $\chi_{xx}$ and $\chi_{yy}$ as functions of the chemical potential $E$ relative to the VBM, revealing that $\chi_{xx}$ and $\chi_{yy}$ are equal, consistent with the symmetry analysis. A significant band splitting is observed near the VBM, as shown in Fig. 4(b). The REE exhibits a substantial magnitude of $1.704 \times 10^{11}\ \frac{\hbar}{e} \frac{1}{\text{eV} \cdot \text{cm}}$ at the energy corresponding to the VBM, where bands with single spin occupancy maximize these two REE tensor components. At lower energies, the REE magnitude decreases due to the emergence of a band with opposite spin, which partially compensates the spin accumulation.\\
%============================================
\section{Representative material B\lowercase{r}G\lowercase{a}T\lowercase{e}}
%============================================
\indent Monolayer BrGaTe possesses orthorhombic lattice symmetry, with its LG and point group being $Pm2a$ (No. 31) and $C_{2v}$, respectively. As depicted in Fig. 5(a), each Ga atom in the middle two layers is bonded to three Te atoms and one Br atom, forming a slightly distorted tetrahedron. Since Te is heavy element, the band structure considering SOC along the high symmetry line is shown in Fig. 5(b). It is observed that this material is a typical semiconductor, with the VBM and CBM located around the $\Gamma$ point. Furthermore, the energy bands of opposite spins are significantly split due to the strong SOC. Considering the wave vector point group $C_{2v}$ and irreducible representation $\Gamma_5$, the effective $\bm{k \cdot p}$ Hamiltonian of lowest-order around VBM can be written as $H^\Gamma_{\text{split}} = k_x\sigma_y$, which normally produce Rashba-Dresselhaus ST\cite{mera2021different}. However, the first-principles calculation shows that its spin orientation is mainly along the z direction, as shown in Fig. 5(c). Due to Rashba-Dresselhaus ST assuming the in-plane spin orientation, it does not apply here. It is more like the persistent spin textures proposed by previous researchers\cite{persist1,persist2}, namely its spin orientation is independent of momentum, but the spin along z direction.\\
\indent Through symmetry analysis of the REE tensor, it is found that the system possesses only a single nonzero tensor component, $\chi_{zx}$. The dependence of the REE magnitude on the chemical potential $E$ is presented in Fig. 5(d). A pronounced peak emerges at $E=-0.631\,\mathrm{eV}$, where the REE value reaches $1.043 \times 10^{11}\,\frac{\hbar}{e}\,\frac{1}{\mathrm{eV} \cdot \mathrm{cm}}$, exceeding previously reported values for 2D materials by an order of magnitude. In this case, the spin polarization points along the $z$ axis, which is analogous to recently reported out-of-plane Edelstein effects\cite{out-Rashba1,out-Rashba2,out-Rashba3}, and we provide a new approach to realizing this effect. Examining the isoenergy contour at $E=-0.631\,\mathrm{eV}$ [see Fig. 5(e)] reveals a shape reminiscent of the conventional Rashba-like pattern: two closed curves shifted laterally in momentum space, except that in the present case, each contour consists of two anisotropic circles. The spin distribution shown in Fig. 5(c) indicates that the spin tends to align along the $z$ direction near the central line of the Fermi surface (apart from the spin-degenerate center line), while it gradually becomes in-plane towards the edges. Consequently, as the hole doping concentration increases, the REE tensor exhibits a characteristic trend of first increasing and then decreasing, as depicted in Fig. 5(d).\\
%============================================
\section{CONCLUSIONS}
%============================================
\indent In summary, we have conducted a systematic symmetry analysis to investigate the REE response tensors across 80 LGs for 2D materials. We began by examining the symmetry constraints imposed by the symmetry operations of each LG on the REE tensor, establishing a clear relationship between these operations and the resulting REE tensor symmetry. This relationship elucidates how the LG of a material dictates the form of its REE tensor. By leveraging the properties of symmetric tensors, we classified the 80 LGs into 13 distinct categories exhibiting specific symmetry constraints. Through a meticulous screening process within the C2DB database, we identified 54 candidate materials with potentially high spin conversion efficiencies. High-precision calculations revealed that these materials possess REE tensors with magnitudes an order of magnitude larger than the maximum values reported in existing literature. The pronounced REE response tensor opens up new opportunities for spin-based logic and spin-orbit torque (SOT) memory technologies. In spin logic architectures, the out-of-plane REE enables the generation of spin-polarized currents that realize reconfigurable logic operations with extremely low energy consumption. In SOT memory devices, the perpendicular spin polarization induces strong damping-like torques, enabling efficient magnetization control. Together, these functionalities establish a foundation for scalable, energy-efficient spintronic platforms suited for next-generation computing and memory technologies.\\
%============================================
\section{METHODS}
%============================================
\indent Density Functional Theory calculations were performed using a plane-wave pseudopotential approach as implemented in the Quantum-ESPRESSO package~\citep{qe1,qe2}. The interactions between ions and electrons were described using fully relativistic projector augmented wave pseudopotentials obtained from the PSLIBRARY database~\citep{pseudopotentials}. The exchange-correlation interactions were modeled within the generalized gradient approximation framework, employing the Perdew-Burke-Ernzerhof functional~\citep{PBE}. A plane-wave energy cutoff of 70~Ry was adopted, and Brillouin zone sampling was conducted using the Monkhorst-Pack scheme, optimized individually for each compound. To ensure precise convergence of the Rashba-Edelstein effect, ultradense k-point grids were employed to interpolate the Hamiltonians using the PAOFLOW code~\citep{paoflow1,paoflow2}. SOC was incorporated self-consistently in all calculations. The detailed calculation parameters were listd in Sec. IV of the SM.\\
%============================================

\bibliography{references}
\end{document}